\documentclass{desyproc}

\begin{document}
\title{Kaons - Recent Results and Future Plans}

\author{{\slshape Takeshi K. Komatsubara}\\[1ex]
KEK, Oho, Tsukuba, Ibaraki 305-0801, Japan }

\contribID{41}

\confID{800}  
\desyproc{DESY-PROC-2009-xx}
\acronym{LP09} 
\doi  

\maketitle

\begin{abstract}
Recent results and future plans of kaon physics are reviewed. 
Topics include 
CP violation, 
rare decays, 
light neutral-boson search, 
lepton flavor universality, 
and CPT and QM tests. 
\end{abstract}

\section{CP violation}

This review article starts with the kaon-side story of CP violation. 
After the $K^0_L\to\pi^+\pi^-$ decay was discovered~\cite{FC} in 1964
and the CP asymmetry in the $K^0-\overline{K^0}$ mixing 
was established~\cite{FC_RMP}, 
a long-standing problem has been its origin; 
the first question was 
whether it was due to the $\Delta S = 2$ {\em superweak} transition~\cite{Wolfenstein} or not. 
In 1973, Kobayashi and Maskawa~\cite{KM} accommodated 
CP violation in the electroweak theory
with six quarks (and single complex-phase). 
The Kobayashi-Maskawa theory~\cite{KM_RMP}, 
including  the prediction of 
{\em direct} CP violation in the decay process 
from the CP-odd component ($K_2$) to the CP-even state ($\pi\pi$), 
was verified by the observations of 
Time-reversal non-invariance (CPLEAR~\cite{CPLEAR} at CERN) and 
finite $\epsilon^{\prime}/\epsilon$ (NA48 at CERN and KTeV at FNAL)
as well as the discoveries of top quark and  CP-violating B decays. 

 The KTeV collaboration reported the final measurement of $Re(\epsilon^{\prime}/\epsilon)$
 with their entire data set: 
      $ ( 19.2 \pm 1.1(stat.) \pm 1.8 (syst.) ) \times 10^{-4}$~\cite{KTeV-final}.
 Combining all the measurements including the final result from NA48:
      $ ( 14.7 \pm 2.2 ) \times 10^{-4}$~\cite{NA48-final}, 
 the world average is
      $ ( 16.8 \pm 1.4 ) \times 10^{-4}$~\cite{BWY_PTP}; 
 it clearly demonstrates the existence of direct CP violation. 
 However, due to theoretical uncertainties in the hadronic matrix elements,  
 to get information on the Standard Model and New Physics beyond it
 from $Re(\epsilon^{\prime}/\epsilon)$ is difficult
 and remains to be a challenge to lattice QCD calculations~\cite{Ukawa}. 
 
 In the modern classification~\cite{BTevRun2} 
 CP violation is grouped into three: 
 in {\em mixing}, {\em decay}, and {\em interference between decays with and without mixing}. 
 All of these have been extensively studied in the B Factory experiments,
 while the study of CP violation in the charged-kaon decay modes started recently. 
 The NA48/2 experiment at CERN performed charge asymmetry measurements
 with the simultaneous $K^{\pm}$ beams of $60\pm 3$ GeV/$c$ in 2003 and 2004.
 The asymmetries of  the linear slope parameter in the matrix element expansion
 of the $K^{\pm} \to \pi^{\pm} \pi^+ \pi^-$ decay and 
 the $K^{\pm} \to \pi^{\pm} \pi^0 \pi^0$ decay
 were measured to be 
      $( -1.5 \pm 2.2 ) \times 10^{-4}$ 
      and
      $( 1.8 \pm 1.8) \times 10^{-4}$
 with the data sets of 4G and 0.1G events, respectively~\cite{NA48Kp3}.
 The Standard Model expectation is in $10^{-5}\sim 10^{-6}$, and 
 no evidence for CP violation in {\em decay} 
 was observed at the level of $2\times 10^{-4}$. 
 NA48/2 also measured the asymmetries of $K^+$ and $K^-$ decay-widths 
 in $K^{\pm} \to \pi^{\pm} e^{+} e^{-}$ and $K^{\pm} \to \pi^{\pm} \pi^0 \gamma$
 to be 
      $ ( -2.2 \pm 1.5(stat.) \pm 0.6 (syst.) ) \times 10^{-2}$~\cite{NA48-piee}
      and
      $ ( 0.0 \pm 1.0(stat.) \pm 0.6 (syst.) ) \times 10^{-3}$~\cite{NA48-ppg}
 and set the upper limits of $2.1\%$ and $0.15\%$, respectively. 

The CP-violating processes in  {\em mixing} and {\em decay} 
suffers from hadronic uncertainties.
In contrast,  
the CP violation in {\em interference between decays with and without mixing}
is theoretically clean,
and the decay $K^0_L\to\pi^0\nu\bar{\nu}$~\cite{litt89} is known to be a golden mode
in this category~\cite{GrossmanNir}
because the branching ratio can be calculated with very small 
theoretical-uncertainties in the Standard Model as well as in New Physics.
Measurement of the branching ratios for $K^0\to\pi^0\nu\bar{\nu}$
and for the charged counterpart $K^+\to\pi^+\nu\bar{\nu}$ is
the main issue 
in the near future plans of kaon physics, and is the topics of the next section.

\section{Rare decays}

\begin{figure}[hb]
\centerline{\includegraphics[width=0.85\textwidth]{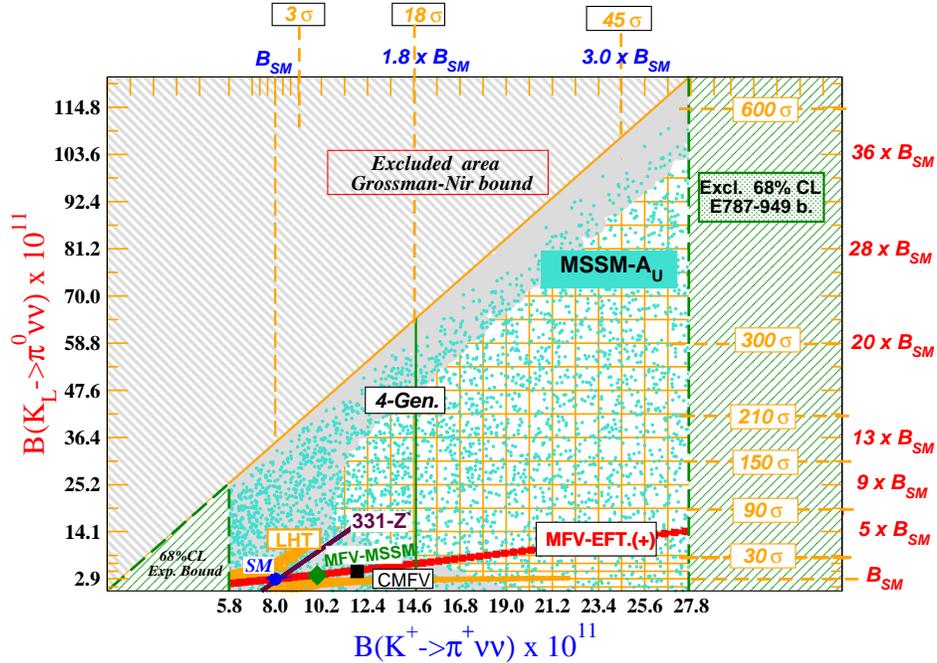}}
\caption{$K\to\pi\nu\bar{\nu}$ branching ratios beyond the Standard Model, 
                by courtesy of F.~Mescia and C.~Smith~\cite{Krarehtml}.}
\label{Fig:Krare_BSM_new}
\end{figure}

The $K\to\pi\nu\bar{\nu}$ decay~\cite{BUS08} 
is a Flavor-Changing Neutral Current (FCNC) process 
and is induced by the electroweak loop effects as 
Penguin and Box diagrams. 
The decay is suppressed in the Standard Model, and
the branching ratios are predicted as
     $B(K^0_L \to \pi^0\nu\bar{\nu}) = ( 2.76 \pm 0.40 ) \times 10^{-11} $
     and
     $B(K^+ \to \pi^+\nu\bar{\nu}(\gamma)) = (8.22 \pm 0.84 ) \times 10^{-11}$~\cite{Krarehtml}, 
in which the uncertainties are dominated 
by the allowed range of the quark-mixing matrix elements.
Long-distance contributions are small,
the hadronic matrix elements are extracted
from the $K^+\to\pi^0 e^+ \nu$ decay~\cite{Mescia-Smith07}, and
the next-to-next-to-leading order QCD corrections~\cite{BGHN06}
and the QED and electroweak corrections~\cite{Brod-Gorbahn08}
to the charm quark contribution to
$K^+\to\pi^+\nu\bar{\nu}$ have been calculated. 
New Physics could affect these branching ratios~\cite{BBIL06} and,  
by the measurement, 
the flavor structure in New Physics (operators and phases in the interactions of new particles) 
can be studied. 
The $K\to\pi\nu\bar{\nu}$ branching ratios beyond the Standard Model
are presented in Fig.~\ref{Fig:Krare_BSM_new}.
A model-independent bound 
$B(K^0_L \to \pi^0\nu\bar{\nu}) < 4.4\times B(K^+ \to \pi^+\nu\bar{\nu})$
(Grossman-Nir bound~\cite{GrossmanNir})
can be extracted from their isospin relation. 

The signature of $K\to\pi\nu\bar{\nu}$ is a kaon decay into {\em a pion plus nothing}.
Background rejection is essential in these experiments, 
and {\em blind analysis} techniques have been developed and refined 
to achieve a high level of confidence in the background measurements.
To verify {\em nothing}, 
hermetic extra-particle detection
by photon and charged-particle detectors, 
called the {\em veto}, 
is imposed to the hits in coincidence with the pion time 
and
with the energy threshold less than a few MeV.
Tight veto requirements are indispensable
in order to achieve a low detection-inefficiency $< 10^{-3}\sim 10^{-4}$; 
good timing resolution for low energy hits is therefore  
essential to avoid acceptance loss due to accidental hits. 

The E391a experiment at KEK was 
the first dedicated search for the $K^0_L \to \pi^0\nu\bar{\nu}$ decay.
A small-diameter neutral beam (called a {\em pencil} beam~\cite{pencil}) 
was developed and constructed. 
The $K^0_L$ beam whose momentum peaked at 2 GeV/$c$ 
was produced by the 12-GeV proton synchrotron (KEK-PS). 
The energy and position of the two photons from $\pi^0$ decays were measured 
by a downstream calorimeter. 
The $K^0_L$-decay vertex position along the beam line was determined
from the constraint of $\pi^0$ mass, 
and a $\pi^0$ with a large transverse momentum ($\ge$ 0.12 GeV/$c$) 
was the signal.
The remaining part of the calorimeter not hit by the two photons 
and the other detector subsystems that covered the decay region
were used as a veto,
which was crucial
to suppress the major background from $K^0_L\to\pi^0\pi^0$.
The beam line and the collimation scheme were designed carefully 
to minimize the beam halo (mostly neutrons), 
which could interact with the counters near the beam 
and produce $\pi^0$'s and $\eta$'s.

\begin{figure}[hb]
\centerline{\includegraphics[width=0.52\textwidth]{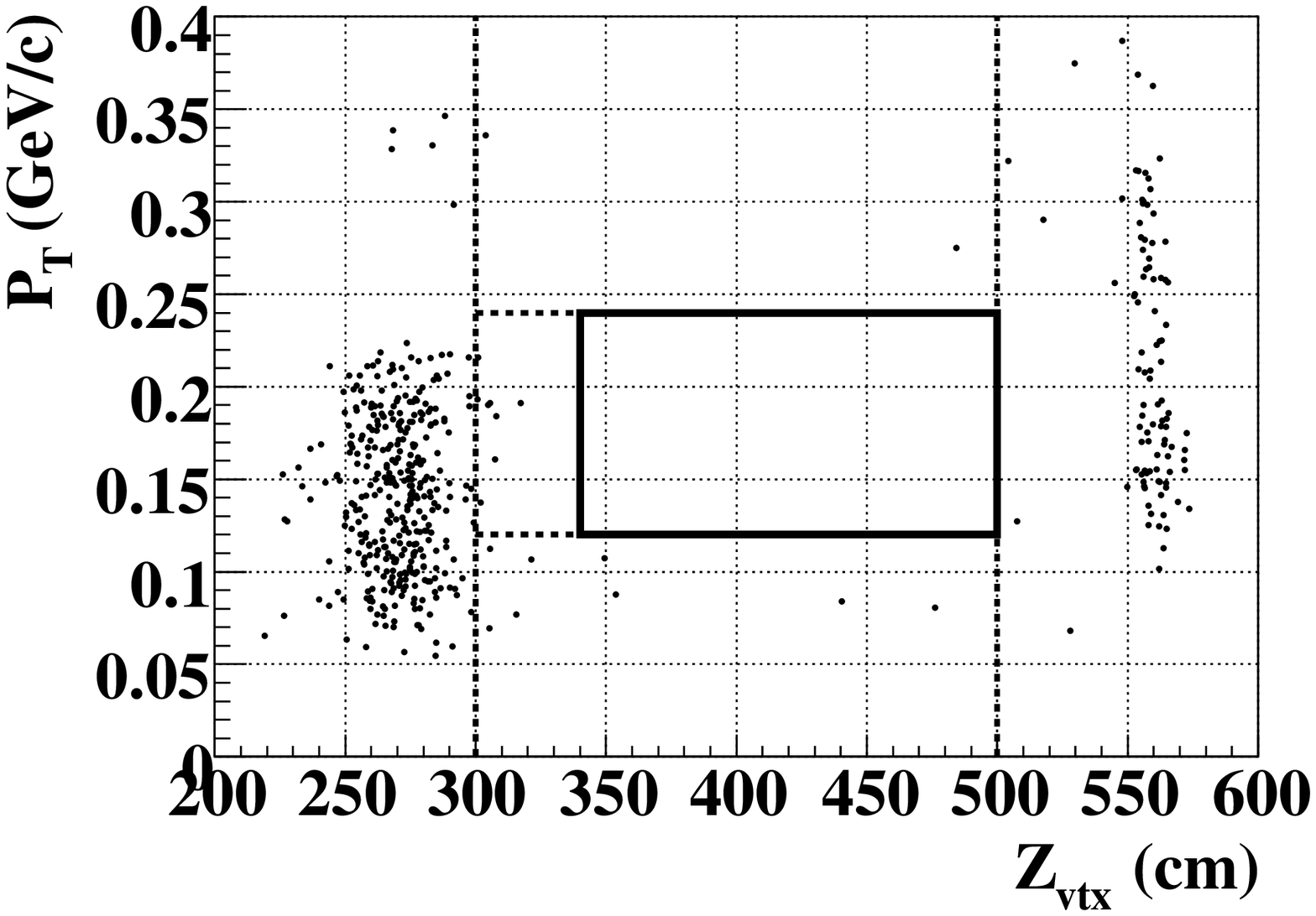}
                    \includegraphics[width=0.45\textwidth]{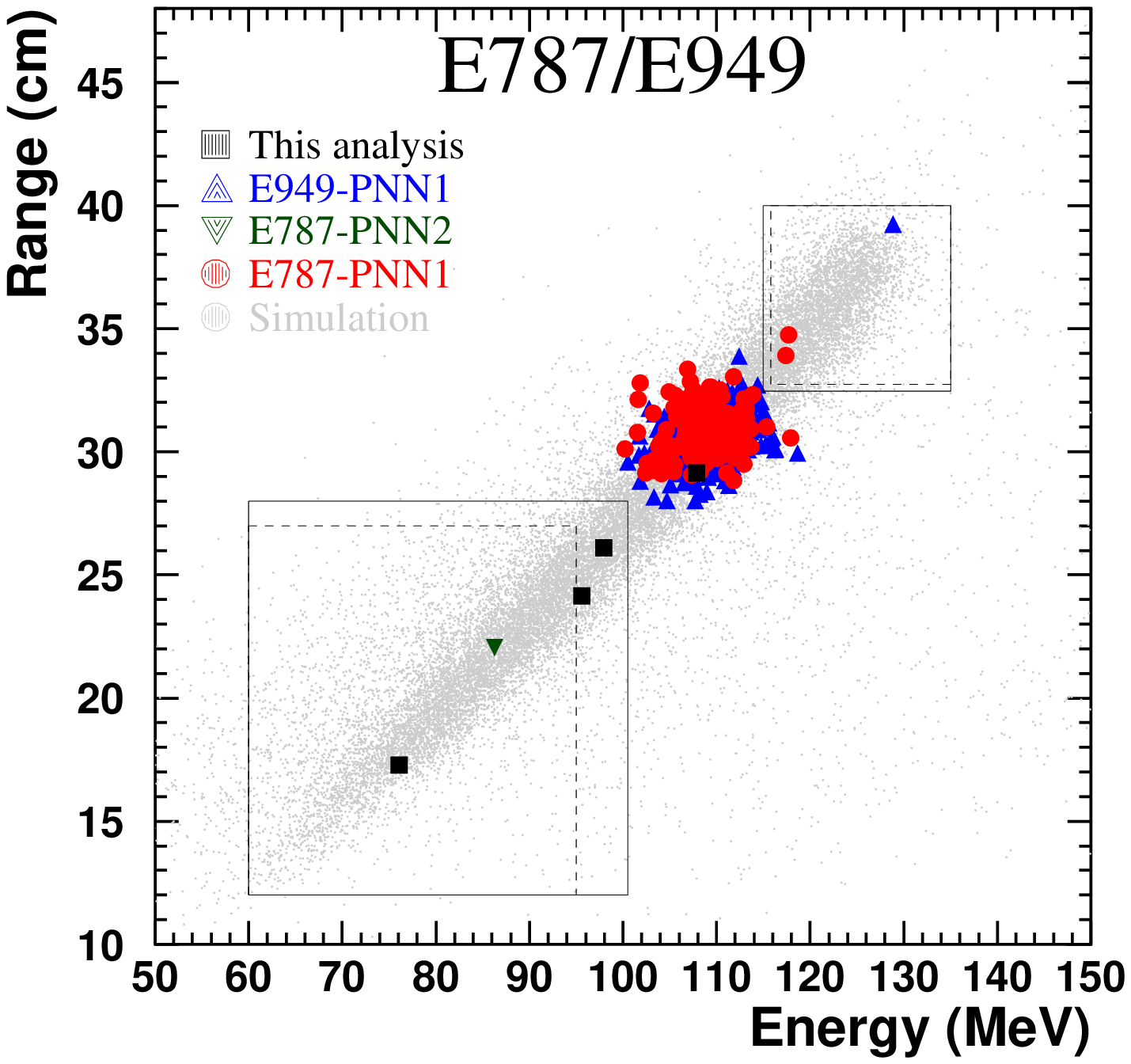}}
\caption{E391a result on $K^0_L \to \pi^0\nu\bar{\nu}$~\cite{E391afinal} (left); 
                E787/E949 result on $K^+ \to \pi^+\nu\bar{\nu}$~\cite{E949final} (right).} 
 \label{Fig:kpnnresults}
\end{figure}

Final results from E391a on $K^0_L \to \pi^0\nu\bar{\nu}$
were published recently~\cite{E391afinal}. 
Combining the data sets from February-April and October-December 2005, 
the single event sensitivity was $1.11\times 10^{-8}$ and no events were observed
inside the signal region (Fig.~\ref{Fig:kpnnresults}, left).
The upper limit on $B(K^0_L \to \pi^0\nu\bar{\nu}) $ was set to be
$2.6\times 10^{-8}$ at the 90\% confidence level (C.L.).
The E391a experiment has improved the limit from previous experiments
by a factor of 20. 

The E949 experiment at BNL measured 
the charged track emanating from $K^+ \to \pi^+\nu\bar{\nu}$
decaying at rest in the stopping target.
Charged-particle detectors for 
measurement of the $\pi^+$ properties
were located in the central region of the detector
and were surrounded by hermetic photon detectors.
The $\pi^+$ momentum ($P_{\pi^+}$) from $K^+ \to \pi^+\nu\bar{\nu}$ is less than 0.227 GeV/$c$, 
while the major background sources of 
$K^+\to\pi^+\pi^0$ and $K^+\to\mu^+\nu$
are two-body decays and 
have monochromatic momentum of 0.205 GeV/$c$ and 0.236 GeV/$c$, respectively.
The $\pi^+$ momentum regions 
above and below the peak from  $K^+\to\pi^+\pi^0$ were adopted.
Redundant kinematic measurement and 
$\mu^+$ rejection were employed; 
the latter was crucial
because the $K^+\to\mu^+\nu$ background had the same topology as the signal. 
Pion contamination to the incident $K^+$ beam  (0.7 GeV/$c$)
was reduced by two stages of electrostatic particle separation in the beam line 
to prevent the background due to scattered beam pions. 

The E949 experiment observed 
one $K^+ \to \pi^+\nu\bar{\nu}$ event in the kinematic region 
$0.211 < P_{\pi^+} < 0.229$ GeV/$c$ (PNN1)~\cite{pnn1final}
and three events in the region $0.140 < P_{\pi^+} < 0.199$ GeV/$c$ (PNN2)~\cite{E949final}.
Combining the results 
with the observation of two events in PNN1 and one event in PNN2
by the predecessor experiment E787 gave 
$B(K^+ \to \pi^+\nu\bar{\nu}) = (1.73^{+1.15}_{-1.05})\times 10^{-10}$
(Fig.~\ref{Fig:kpnnresults}, right)~\cite{E949final}, 
consistent with the Standard Model prediction.

The next generation of $K^0_L \to \pi^0\nu\bar{\nu}$
is the E14 KOTO experiment~\cite{KOTO} 
at the new high-intensity proton accelerator facility 
J-PARC (Japan Proton Accelerator Research Complex)~\cite{JPARC}. 
The accelerators, consisting of a Linac, 3-GeV Rapid Cycle Synchrotron and Main Ring,  
succeeded in the acceleration to 30 GeV and slow and fast beam-extractions.
The KOTO collaboration built the neutral beam line  at the Hadron Hall of J-PARC
and surveyed the beam in 2009. 
They started the detector construction in 2010
with the undoped CsI crystals used in the KTeV calorimeter.
The next generation of $K^+ \to \pi^+\nu\bar{\nu}$ 
is the NA62 experiment~\cite{NA62} at CERN,
which will use $K^+$ decays in flight
from an un-separated beam of 75 GeV/$c$.
The detector R\&D with beam tests is close to the end, 
and the NA62 detector is being built. 
KOTO, as the first step in measuring $B(K^0_L \to \pi^0\nu\bar{\nu})$ at J-PARC, 
aims at the observation of $K^0_L \to \pi^0\nu\bar{\nu}$,
and the goal of NA62 is to detect 100 $K^+ \to \pi^+\nu\bar{\nu}$ events.
At FNAL, a new proposal~\cite{FNALP996} 
to measure $K^+ \to \pi^+\nu\bar{\nu}$ decays at rest 
has been submitted.
Higher sensitivity kaon experiments based on a new high-intensity proton source
at FNAL~\cite{ProjectX} are now under discussion. 

At J-PARC, 
another new kaon experiment (E06 TREK)~\cite{TREK} is being prepared. 
In the $K^+\to\pi^0\mu^+\nu$ decay, 
the transverse muon polarization ($P_T$)
is a T-odd quantity and is a CP violation observable. 
New sources of CP violation may give rise to $P_T$ as large as $10^{-3}$.
TREK is a successor to E246 at KEK-PS~\cite{E246final} 
and will measure  the charged track and photons from the $K^+$ decay at rest
with the E246 superconducting toroidal magnet,
and aims at a $P_T$ sensitivity of $10^{-4}$.
A low-momentum beam line is being built at the Hadron Hall.

\section{Light neutral-boson search}

Experimental searches for very light bosons have a long history, but
a neutral boson whose mass is twice the muon mass has not yet been 
excluded.
In 2005, the HyperCP collaboration at FNAL reported
three events of the $\Sigma^+\to p \mu^+ \mu^-$ decay, 
and the dimuon mass may indicate a neutral intermediate state $P^0$ 
with a mass of $214.3\pm 0.5$ MeV/$c^2$~\cite{HyperCP}.
Since the events were observed in an FCNC with a strange to down quark transition,
$P^0$ should be confirmable with kaon decays. 
Dimuon masses in previous $K^{+}\to\pi^{+}\mu^+\mu^-$  measurements
were not observed in the narrow range around the $P^0$ mass; 
thus, $P^0$ should be a pseudo-scalar or axial-vector particle 
and be studied with the three-body decay $K\to\pi\pi P^0$.

The KTeV collaboration searched for the $K^0_L \to \pi^0\pi^0\mu^+\mu^-$ decay
for the first time~\cite{KTeV-ppmm} and set 
$B(K^0_L \to \pi^0\pi^0\mu^+\mu^-) < 8.63\times 10^{-11} $
and 
$B(K^0_L \to \pi^0\pi^0 P^0\to\pi^0\pi^0\mu^+\mu^-) < 9.44\times 10^{-11}$
(90\% C.L.).
The E391a collaboration searched 
for the decay $K^0_L \to \pi^0\pi^0 X$, $X\to\gamma\gamma$ 
in the mass range of $X$ from 194.3 to 219.3 MeV/$c^2$, 
and set 
$B(K^0_L \to \pi^0\pi^0 P^0\to\pi^0\pi^0\gamma\gamma) < 2.4\times 10^{-7}$
(90\% C.L.)~\cite{E391a-ppgg}.
Both results almost ruled out the predictions 
when $P^0$ is a pseudo-scalar particle~\cite{HTV07,OT09}. 

The E787/E949 results on $K^+ \to \pi^+\nu\bar{\nu}$
have also been interpreted 
in the two-body decay $K^+\to\pi^+ X$, 
where $X$ is a massive noninteracting particle either in stable or unstable, 
and in $K^+ \to \pi^+ P^0$, $P^0\to\nu\bar{\nu}$.
The limits are presented in~\cite{E949final}.

\section{Lepton flavor universality}

Investigation of  
the lepton-flavor violating (LFV) processes involving both quarks and charged leptons
($K^0_L\to \mu^{\pm} e^{\mp}$ by E871 at BNL~\cite{E871},
 $K^{+}\to\pi^{+} \mu^{+} e^{-}$, 
 $K^{+}\to\pi^{-} \mu^{+} \mu^{+}$, and
 $K^{+}\to\pi^{-} e^{+} e^{+}$
 by E865 at BNL~\cite{E865,E865exotic}, 
$K^0_L\to \pi^0 \mu^{\pm} e^{\mp}$ and $K^0_L\to \pi^0 \pi^0 \mu^{\pm} e^{\mp}$ 
by KTeV~\cite{KTeVLFV})
has achieved stringent limits on the branching ratios in  $10^{-9}\sim 10^{-12}$. 
Continual efforts are made to search for the $\mu^+\to e^+\gamma$ decay
and the $\mu^{-}N\to e^{-}N$ conversion with muons~\cite{Mori}.

\begin{figure}[hb]
\centerline{\includegraphics[width=0.45\textwidth]{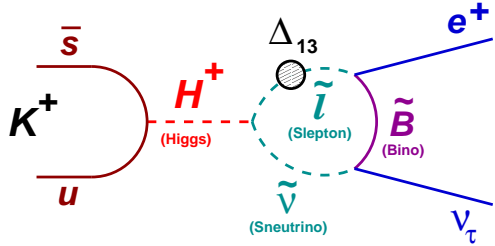}}
\caption{Diagram for the $K^+\to e^+ \nu_{\tau}$ decay.}
 \label{Fig:LFUdiagram}
\end{figure}

The LFV process in kaon decays is currently studied, intensively,
in the context of high precision tests of Lepton Flavor universality. 
The ratio $R_{K}
\equiv \Gamma(K^+\to\ e^+\nu(\gamma)) / \Gamma(K^+\to\ \mu^+\nu(\gamma))$
is helicity suppressed in the Standard Model
due to the V-A couplings
and is predicted to be 
$R_{K}^{SM} = (2.477\pm 0.001)\times 10^{-5}$~\cite{CiriglianoRosell}, 
in which the radiative decay $K^+\to e^+ \nu\gamma$ ($K_{e2\gamma}$) via
internal bremsstrahlung  is included. 
Suppose a decay $K^+\to e^+ \nu_{\tau}$ exists due to the process of 
an intermediate charged-Higgs particle and a LFV Supersymmetric loop
(Fig.~\ref{Fig:LFUdiagram})~\cite{MPP06}.
Since the neutrino flavor is undetermined experimentally,
deviations of 
\[  
R_{K}\ =\ 
\frac{\Sigma_i \ \Gamma(K^+\to\ e^+\nu_i)}
        {\Sigma_i \ \Gamma(K^+\to\ \mu^+\nu_i)}
\ \simeq\ 
  \frac{\Gamma_{SM}(K^+\to\ e^+\nu_{e}) + \Gamma_{NP}(K^+\to\ e^+\nu_{\tau}) } 
          {\Gamma_{SM}(K^+\to\ \mu^+\nu_{\mu})}
\]
from the Standard Model prediction, $\Delta R_{K}$, 
in the relative size of $10^{-2}\sim 10^{-3}$ are suggested~\cite{MPP08}
as a function of the charged Higgs mass $m_{H^{+}}$, 
the ratio of the Higgs vacuum expectation values for the up- and down- quark masses
(denoted as $\tan \beta$), 
and the effective $e-\tau$ coupling constant $\Delta_{13}$:
\[
   \frac{\Delta R_{K}}{R_{K}^{SM}}\ =\
      \frac{\Gamma_{NP}(K^+\to\ e^+\nu_{\tau})}{\Gamma_{SM}(K^+\to\ e^+\nu_{e})}\ =\ 
   \left( \frac{m_{K^{+}}}{m_{H^{+}}}   \right)^4\ 
   \left( \frac{m_{\tau}}{m_{e}} \right)^2\
   | \Delta_{13} |^2\
   \tan^6 \beta 
\]
and can be experimentally studied.
The same physics goal is pursued 
by the PIENU experiment~\cite{PIENU} at TRIUMF
and the PEN experiment~\cite{PEN} at PSI
to measure 
$\Gamma(\pi^+\to\ e^+\nu(\gamma)) / \Gamma(\pi^+\to\ \mu^+\nu(\gamma))$
precisely.

\begin{figure}[hb]
\centerline{\includegraphics[width=0.45\textwidth]{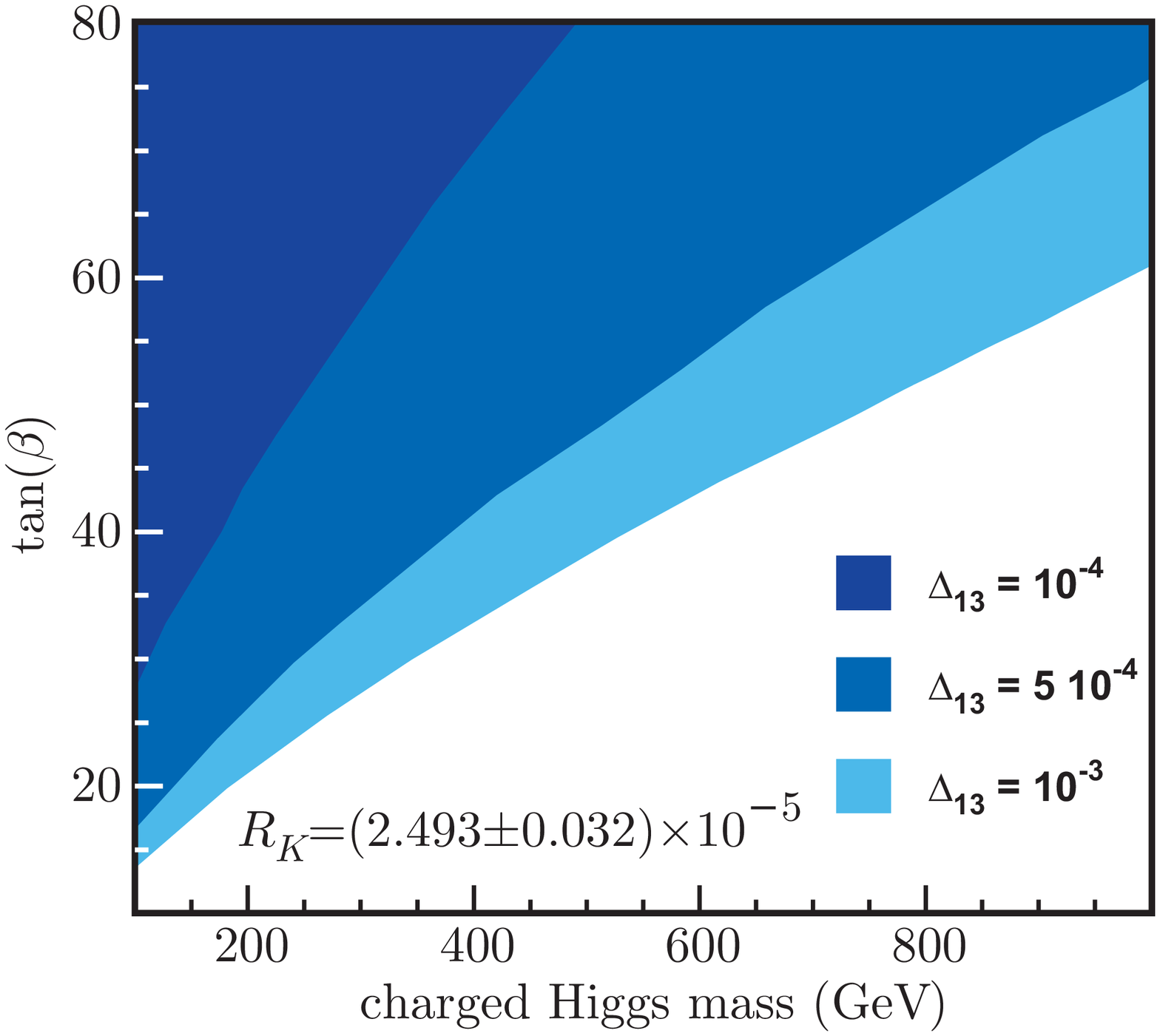}
                    \includegraphics[width=0.52\textwidth]{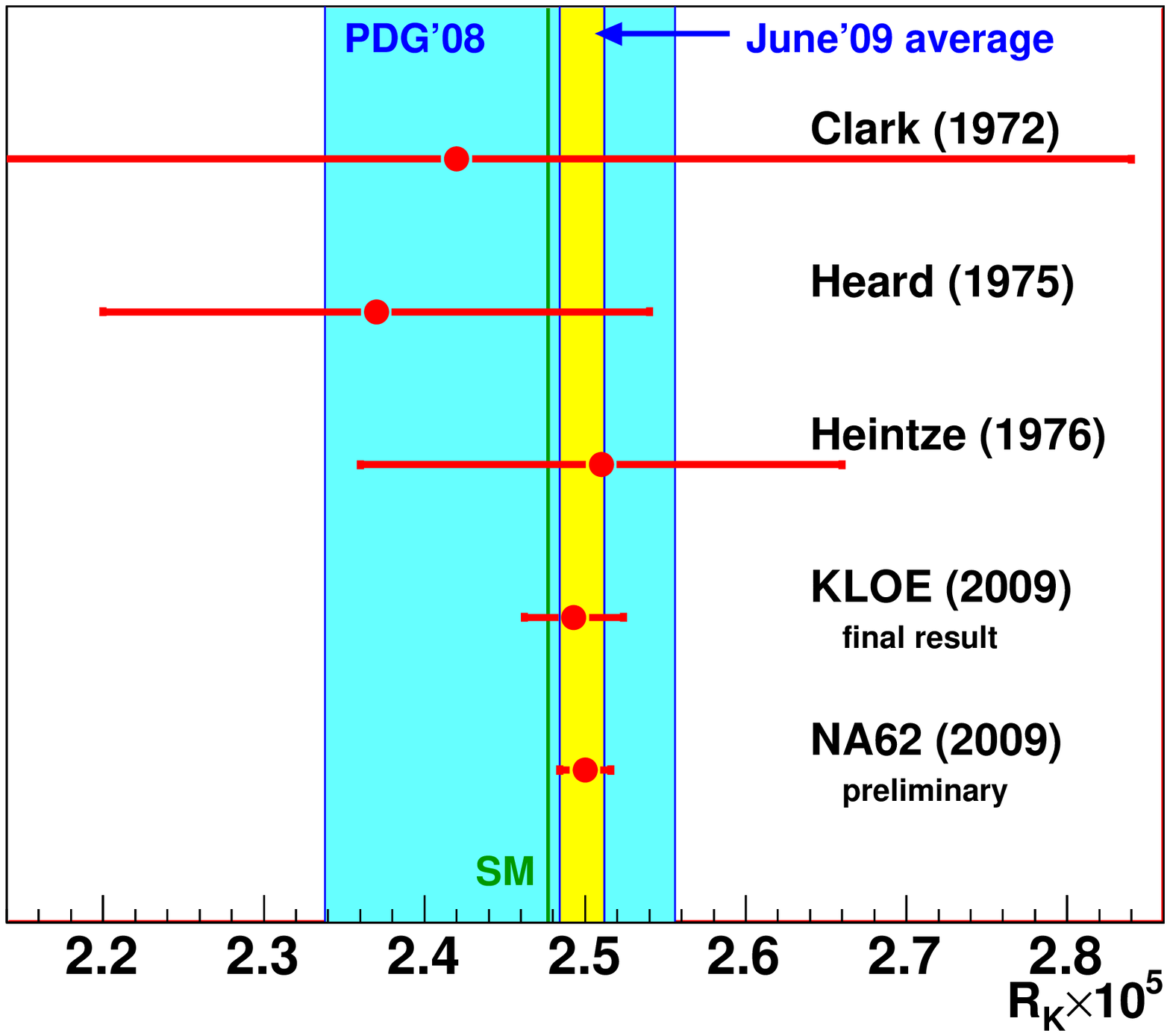}}
\caption{KLOE result on the excluded regions at 95\% C.L. in the plane $m_{H^{+}}-\tan \beta$
                for $\Delta_{13}$=$10^{-4}$, $5\times 10^{-4}$ and $10^{-3}$~\cite{KLOE-RK} (left); 
                summary of the $R_K$ measurements including the report from NA62~\cite{NA62-RK} (right).} 
 \label{Fig:LFUresults}
\end{figure}

The KLOE collaboration at DA$\Phi$NE, the Frascati $\phi$ factory,
measured $R_{K}$ 
with 3.3G 
of the $K^+K^-$ pairs from $\phi$ mesons
with an integrated luminosity of 2.2 fb$^{-1}$ collected during 2001-2005~\cite{KLOE-RK}.
The $K^{\pm} \to\ \ell^{\pm}\nu$ decay in flight ($\sim$0.1 GeV/$c$) 
was reconstructed 
by the tracks of a kaon and a decay product with the same charge 
in the cylindrical drift chamber, and
the squared mass $m_{\ell}^2$
of the lepton for the decay was computed.
To distinguish the $K^{\pm} \to\ e^{\pm}\nu$ events around $m_{\ell}^2=0$
from the tail of the $K^{\pm} \to\ \mu^{\pm}\nu$ peak,
in addition to the track quality cuts, 
information about shower profile and total energy deposition 
in the electromagnetic calorimeter, combined with a neural network,
and time-of-flight information were used for electron identification.
The numbers of $K\to e\nu(\gamma)$ events were 
$7064\pm 102$ for $K^+$ and
$6750\pm 101$ for $K^-$, respectively, 
89.8\% of which were  $K\to e\nu$ events and 
the $K_{e2\gamma}$ events with $E_{\gamma}<10$ MeV.
The contribution from the $K_{e2\gamma}$ events with $E_{\gamma}>10$ MeV, 
due to the direct-emission process, 
was studied~\cite{KLOE-RK,KLOE-rad} by using a separate sample 
with photon detection requirement, 
and was subtracted.
The numbers of $K\to \mu\nu(\gamma)$ events were 
287.8M for $K^+$ and
274.2M for $K^-$, respectively.
The difference between the $K^+$ and $K^-$ counts was
due to the larger cross section of $K^{-}$ nuclear interaction 
in the material traversed.
Finally, KLOE obtained 
$R_{K}\ =\  ( 2.493 \pm 0.025(stat.) \pm 0.019 (syst.) ) \times 10^{-5}$,
in agreement with the Standard Model prediction. 
The regions excluded at 95\% C.L. in the plane  $M_{H^{+}}-\tan \beta$
are shown in Fig.~\ref{Fig:LFUresults}, left, 
for different values of $\Delta_{13}$.

The NA62 collaboration, as the first phase, 
collected data to measure $R_{K}$ during four months in 2007, 
and collected special data samples to study systematic effects for two weeks in 2008.
The beam line and detector apparatus of NA48/2 were used; 
90\% of the data were taken with the $K^+$ beam of $74.0\pm 1.6$ GeV/$c$,
because the muon sweeping system provided better suppression 
of the positrons produced by beam halo muons via $\mu\to e$ decay.
Preliminary results based on the analysis of 40\% of the 2007 data 
collected with the $K^+$ beam only
were reported in~\cite{NA62-RK}.
The number of $K^+\to e^+\nu$ candidate events was $51089$,
and the number of $K^+\to \mu^+\nu$ candidate events was
15.56M.
The source of the  main background, $(6.28\pm 0.17)$ \%,
was found to be the $K^+ \to \mu^+ \nu$ decay 
with muon identification as positron 
due to {\em catastrophic} bremsstrahlung
in or in front of the liquid-Krypton electromagnetic calorimeter (LKr).
The probability of the mis-identification
was studied with pure muon samples, 
without positron contamination due to $\mu^+\to e^+$ decays in flight, 
selected from the tracks traversing a $9.2X_0$-thick lead wall installed in front of LKr. 
$R_{K}$ was obtained to be 
$( 2.500 \pm 0.012(stat.) \pm 0.011 (syst.) ) \times 10^{-5}$,
consistent with the Standard Model  prediction.
Combining with other $R_K$ measurements, 
the current world average is
$(2.498\pm 0.014) \times 10^{-5}$ as presented in Fig.~\ref{Fig:LFUresults}, right. 
The final results from this 40\% partial data will be available soon.

The $R_K$ measurement will continue 
in the NA62 analysis of the full data sample
as well as in the future KLOE-2 experiment, 
which is described in the next section.

\section{CPT and QM tests}

$\phi$ mesons decay into $K^+ K^-$ pairs with 49\% and $K^0_L K^0_S$ pairs with 34\%. 
In the latter, the initial state is a coherent (and entangled) quantum state:
\[
 |i\rangle\ 
      =\  \frac{1}{\sqrt{2}}\ [\  |K^0\rangle |\overline{K^0}\rangle\ -\ |\overline{K^0}\rangle |K^0\rangle \ ]\  
      =\  \frac{N}{\sqrt{2}}\ [\  |K^0_S\rangle |K^0_L\rangle\ -\ |K^0_L\rangle |K^0_S\rangle \ ]
\]
where $N\simeq 1$ is a normalization factor. 
In the KLOE experiment, 
by tagging  $K^0_L$ {\em crash} events in the calorimeter, 
a pure $K^0_S$ beam was available and 
there has been a major improvement in $K^0_S$ decay measurements~\cite{KLOEreview}. 
With the results of various new measurements on neutral kaon decays,
the Bell-Steinberger relation was used~\cite{KLOE-BS} 
to provide a constraint 
relating the unitarity of the sum of the decay amplitudes to the CPT observables.
The latest limit on the mass difference between $K^0$ and $\overline{K^0}$ 
was $4.0\times 10^{-19}$ GeV at 95\% C.L.~\cite{KLOE-CPT}.

\begin{figure}[hb]
\centerline{\includegraphics[width=0.45\textwidth]{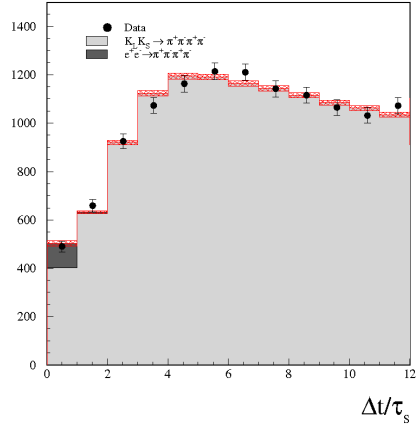}}
\caption{KLOE result on the fit to the measured $I(\Delta t)$ distribution 
               of $\phi \to K^0_S K^0_L\to \pi^+\pi^-\pi^+\pi^-$~\cite{KLOE-dec}.
               The black points with errors are data and the solid histogram is the fit result. 
               The uncertainty arising from the efficiency correction is shown as the hatched area.}
\label{Fig:QMdist}
\end{figure}

In the CP-violating process 
$\phi \to K^0_S K^0_L\to \pi^+\pi^-\pi^+\pi^-$, 
KLOE observed the quantum interference between two kaons for the first time~\cite{KLOE-int}. 
The measured $\Delta t$ distribution,
with $\Delta t$ the absolute value of the time difference of the two $\pi^+\pi^-$ decays, 
can be fitted with the distribution in the $K^0_S-K^0_L$ basis: 
\begin{eqnarray*}
I(\Delta t) & \propto & 
      e^{- \Gamma_L \Delta t}\ +\ e^{- \Gamma_S \Delta t}\ 
        -\ 2(1-\zeta_{SL})\ e^{-  \frac{(\Gamma_S+\Gamma_L)}{2} \Delta t}\ \cos (\Delta m \Delta t)   \\
                 & \rightarrow &
            2 \zeta_{SL} \ \left( 1-  \frac{(\Gamma_S+\Gamma_L)}{2} \Delta t \right)
                      \ \ \ \ \ \ \ \ \ \ \Delta t\rightarrow 0
\end{eqnarray*}
where $\Delta m$ is the mass difference between $K^0_L$ and $K^0_S$.
The interference term 
$e^{-  \frac{(\Gamma_S+\Gamma_L)}{2} \Delta t}\ \cos (\Delta m \Delta t)$
is multiplied by a factor $(1-\zeta_{SL})$
with a {\em decoherence} parameter $\zeta_{SL}$, which represents  
a loss of coherence during the time evolution of the states
and should be zero in Quantum Mechanics (QM).
Final results obtained from KLOE
with 1.5 fb$^{-1}$ in 2004-2005 were~\cite{KLOE-dec}
$\zeta_{SL}\ =\ ( 0.3 \pm 1.8(stat.) \pm 0.6 (syst.) ) \times 10^{-2}$
(Fig.~\ref{Fig:QMdist}) and,
to the fit with the distribution in the $K^0-\overline{K^0}$ basis,
$\zeta_{0\overline{0}}\ =\ ( 1.4 \pm 9.5(stat.) \pm 3.8 (syst.) ) \times 10^{-7}$;
no deviation from QM was observed.
Decoherence in the $K^0-\overline{K^0}$ basis
results in the CP-allowed 
$K^0_S K^0_S\to \pi^+\pi^-\pi^+\pi^-$ decays and thus
the value for the decoherence parameter $\zeta_{0\overline{0}}$ is much smaller.
Since the measurement of non-zero $\zeta_{SL}$ is sensitive to the distribution
in the small $\Delta t$ region, 
the decay vertex resolution due to charged-track extrapolation
($\sim 1\tau_{S}\simeq 6$mm,   
 to the vertex of a $K^0_S\to\pi^+\pi^-$ decay close to the interaction point, in the KLOE detector)
should be improved
in future experiments. 
Other tests of CPT invariance and the basic principles of QM 
are discussed in~\cite{KLOE-dec,CPTQM}.

Kaon physics at the $\phi$ factory will continue 
with an upgraded KLOE detector, KLOE-2~\cite{KLOE-2}, 
at an upgraded DA$\Phi$NE $e^+e^-$ collider. 
During 2008
a new interaction scheme ({\em Crabbed Waist} collisions)  was tested 
with the goal of reaching a peak luminosity of $5\times 10^{32}$ cm$^{-2}$s$^{-1}$,
a factor of three larger than what previously obtained. 
In the first phase starting in 2010, the detector with minimal upgrade
(two devices along the beam line to tag the scattered electrons/positrons 
 from $\gamma\gamma$ collisions) 
restarts taking data for the integrated luminosity of 5 fb$^{-1}$.
To the next phase, 
with the detector upgrades being planned for late 2011, 
an integrated luminosity of 20 fb$^{-1}$ is expected. 
A cylindrical GEM detector~\cite{KLOE2-GEM}
will be placed between the beam pipe and the inner wall of the drift chamber,  
as a new Inner Tracker, 
to improve the decay vertex resolution
and to increase the acceptance for low transverse-momentum tracks.

\section{Conclusions}

The study of kaon physics continues to make great strides. 
The current program to study CP violation is being completed; 
the CP asymmetries in charged kaon decays have not been observed yet. 
The rare decays $K^0_L \to \pi^0\nu\bar{\nu}$, $K^+ \to \pi^+\nu\bar{\nu}$ 
and the lepton flavor universality in 
$\Gamma(K^+\to\ e^+\nu(\gamma)) / \Gamma(K^+\to\ \mu^+\nu(\gamma))$
will be measured 
with highly sophisticated detectors. 
A light neutral boson as a scalar/vector/pseudo-scalar particle was almost ruled out. 
CPT and QM tests will continue in a $\phi$ factory experiment. 
The kaon experiments, with ultra-high sensitivities and precisions, 
are essential and crucial 
as a probe of New Physics beyond the Standard Model.

Important topics of kaon physics:
New Physics effects in $\epsilon_{K}$ and 
other meson-antimeson mixing observables~\cite{Lunghi-Soni08,Buras-Guadagnoli08}, 
$V_{us}$ measurement with kaons and CKM unitarity test~\cite{Prell}
(and the activities in the FlaviaNet Kaon Working Group~\cite{FlaviaNet,FN-Vus}),
basic observables such as $\Delta m$, lifetimes, $\eta_{+-}$, and absolute branching ratios, 
radiative kaon decays, 
hadronic matrix elements and form factors, and
$\pi\pi$ scattering (and the {\em cusp} effect)
are not covered.
The web site of the 2009 Kaon International Conference (KAON09)~\cite{KAON09}
and the Proceedings~\cite{KAON09Proc}
would be considered as a companion to this review.

\section*{Acknowledgments}

I would like to thank
     E.~Blucher,
     P.~Branchini,
     D.A.~Bryman,
     A.J.~Buras, 
     A.~Ceccucci,
     E.~De Lucia,
     A.~Di Domenico,
     E.~Goudzovski, 
     J.~Imazato,
     T.~Inagaki, 
     D.E.~Jaffe, 
     L.S.~Littenberg,
     F.~Mescia,
     M.~Moulson, 
     H.~Morii,
     D.G.~Phillips, 
     G. Ruggiero,
     B.~Sciascia,  
     C.~Smith,
     R.~Tschirhart,
     S.H.~Kettell, 
     R.~Wanke, 
and
     T.~Yamanaka
for providing me help with my talk and this Proceedings article 
for the Lepton Photon 2009 conference.
I would like to acknowledge support from 
Grant-in-Aid for Scientific Research on Priority Areas:
"New Developments of Flavor Physics" by the MEXT Ministry of Japan.


\begin{footnotesize}


\end{footnotesize}



\begin{thebibliography}{99}
\bibitem{FC} 
    J.H.~Christenson, J.W.~Cronin, V.L.~Fitch, and R.~Turlay, 
        Phys. Rev. Lett. {\bf 13} 138 (1964).
\bibitem{FC_RMP} 
    V.L.~Fitch, 
        Rev. Mod. Phys. {\bf 53} 367 (1981);
    J.W.~Cronin, 
        Rev. Mod. Phys. {\bf 53} 373 (1981). 
\bibitem{Wolfenstein} 
    L.~Wolfenstein, 
        Phys. Rev. Lett. {\bf 13} 562 (1964).
\bibitem{KM}
    M.~Kobayashi and T.~Maskawa, 
        Prog. Theor. Phys. {\bf 49}  652 (1973). 
\bibitem{KM_RMP} 
    M.~Kobayashi, 
        Rev. Mod. Phys. {\bf 81} 1019 (2009);  
    T.~Maskawa, 
        Rev. Mod. Phys. {\bf 81} 1027 (2009).
\bibitem{CPLEAR}
    CPLEAR collaboration, A.~Angelopoulos {\it et al.}, 
        Phys. Rep. {\bf 374} 165 (2003),
        and references therein.  
\bibitem{KTeV-final}
    E.T.~Worcester, 
        arXiv:0909.2555 (2009) and arXiv:0910.3160 (2009).
\bibitem{NA48-final}
    J. R.~Batley {\it et al.}, 
        Phys. Lett. {\bf B544} 97 (2002).
\bibitem{BWY_PTP}
    E.g., E.~Blucher, B.~Winstein, and T.~Yamanaka,  
        Prog. Theor. Phys. {\bf 122} 81 (2009). 
\bibitem{Ukawa}
    E.g., A.~Ukawa, 
        in these Proceedings. 
\bibitem{BTevRun2}
    E.g., K.~Anikeev {\it et al.},
    {\it B Physics at the Tevatron: Run II and Beyond}, 
        arXiv:hep-ph/0201071 (2002).
\bibitem{NA48Kp3}
    NA48/2 Collaboration, J. R.~Batley {\it et al.},
        Eur. Phys. J. {\bf C52} 875 (2007).
\bibitem{NA48-piee}
    NA48/2 Collaboration, J.R.~Batley {\it et al.},
         Phys. Lett. {\bf B677} 246 (2009).
\bibitem{NA48-ppg}
    R.~Wanke,
        PoS KAON09 037 (2009);
        NA48/2 Collaboration, J.R.~Batley {\it et al.},
        arXiv:1004.0494 (2010).
\bibitem{litt89}
    L.S.~Littenberg,
        Phys. Rev. {\bf D39} 3322 (1989).
\bibitem{GrossmanNir}
    Y.~Grossman and Y.~Nir,
         Phys. Lett. {\bf B398} 163 (1997).
\bibitem{BUS08} 
    A.J.~Buras, S.~Uhlig, and F.~Schwab,
        Rev. Mod. Phys. {\bf 80} 965 (2008), and references therein.   
\bibitem{Krarehtml}
    F.~Mescia and C.~Smith, 
    {\it $K\to \pi \nu\nu$ decay in the Standard Model},\\
        {\tt   http://www.lnf.infn.it/wg/vus/content/Krare.html} (2010), and references therein.
\bibitem{Mescia-Smith07}
    F.~Mescia and C.~Smith, 
         Phys. Rev. {\bf D76} 034017 (2007).
\bibitem{BGHN06}
    A.J.~Buras, M.~Gorbahn, U.~Haisch, and U.~Nierste,
           JHEP {\bf 0611} 002 (2006).
\bibitem{Brod-Gorbahn08} 
    J.~Brod and M.~Gorbahn,   
          Phys. Rev. {\bf D78} 034006 (2008).
\bibitem{BBIL06}
    D.~Bryman, A.J.~Buras, G.~Isidori, and L.~Littenberg, 
          Int. J. Mod. Phys. {\bf A21}  487 (2006).
\bibitem{pencil}
    H.~Watanabe {\it et al.},
        Nucl. Instrum. Methods. Phys. Res. Sect. A {\bf 545} 542 (2005).
\bibitem{E391afinal}
    E391a Collaboration, J.K.~Ahn {\it et al.},
         Phys. Rev. {\bf D81} 072004 (2010).
\bibitem{pnn1final}
    S.~Adler {\it et al.},
         Phys. Rev. {\bf D77} 052003 (2008).
\bibitem{E949final}
    E949 Collaboration, A.V.~Artamonov {\it et al.},
    Phys. Rev. Lett. {\bf 101} 191802 (2008) 
    and
    Phys. Rev. {\bf D79} 092004 (2009).
\bibitem{KOTO}
        {\tt   http://koto.kek.jp/}; 
         H.~Nanjo, 
        PoS KAON09 047 (2009).
\bibitem{JPARC}
        {\tt   http://j-parc.jp/}.
\bibitem{NA62}
        {\tt   http://na62.web.cern.ch/NA62/}; 
         G.~Ruggiero,
        PoS KAON09 043 (2009).
\bibitem{FNALP996}
       Fermilab Proposal P996: 
       {\it Measurement of the $K^+ \to \pi^+\nu\bar{\nu}$ Decay at Fermilab} (2009).
\bibitem{ProjectX}
       {\tt http://projectx.fnal.gov/}. 
\bibitem{TREK}
        {\tt   http://trek.kek.jp/}; 
         J.~Imazato, 
        PoS KAON09 007 (2009).
\bibitem{E246final}
    E246 Collaboration, M.~Abe {\it et al.},
         Phys. Rev. {\bf D73} 072005 (2006).
\bibitem{HyperCP}
    HyperCP Collaboration, H.K.~Park {\it et al.},
    Phys. Rev. Lett. {\bf 94} 021801 (2005). 
\bibitem{KTeV-ppmm}
    D.G.~Phillips, 
        PoS KAON09 039 (2009).
\bibitem{E391a-ppgg}
    E391a Collaboration, Y.C.~Tung {\it et al.},
    Phys. Rev. Lett. {\bf 102} 051802 (2009). 
\bibitem{HTV07}
    X.G.~He, J.~Tandean, and G.~Valencia, 
    Phys. Rev. Lett. {\bf 98} 081802 (2007). 
\bibitem{OT09}
    S.~Oh and J.~Tandean,
         JHEP {\bf 1001} 022 (2010),
         and references therein. 
\bibitem{E871}
    BNL E871 Collaboration, D.~Ambrose {\it et al.},
    Phys. Rev. Lett. {\bf 81} 5734 (1998). 
\bibitem{E865}
    A.~Sher {\it et al.},
         Phys. Rev. {\bf D72} 012005 (2005).
\bibitem{E865exotic}
    R.~Appel {\it et al.},
         Phys. Rev. Lett. {\bf 85} 2877 (2000).
\bibitem{KTeVLFV}
    KTeV Collaboration, E.~Abouzaid {\it et al.},
    Phys. Rev. Lett. {\bf 100} 131803 (2008). 
\bibitem{Mori}
    T.~Mori, 
        in these Proceedings. 
\bibitem{CiriglianoRosell}
        V.~Cirigliano and I.~Rosell,
        Phys. Rev. Lett. {\bf 99} 231801 (2007). 
\bibitem{MPP06}
    A.~Masiero,  P.~Paradisi, and R.~Petronzio, 
    Phys. Rev. {\bf D74} 011701(R) (2006). 
\bibitem{MPP08}
    A.~Masiero,  P.~Paradisi, and R.~Petronzio, 
    JHEP {\bf 0811} 042 (2008).
\bibitem{PIENU}
     {\tt  http://pienu.triumf.ca/}.
\bibitem{PEN}
     {\tt  http://pen.phys.virginia.edu/}.
\bibitem{KLOE-RK}
   KLOE Collaboration, F.~Ambrosino {\it et al.},
   Eur. Phys. J. {\bf C64} 627 (2009) and {\bf C65} 703(E) (2010). 
\bibitem{KLOE-rad}
   M.~Moulson, 
        PoS KAON09 035 (2009).
\bibitem{NA62-RK}
         E.~Gooudzovski, 
        PoS KAON09 025 (2009) and arXiv:1005.1192 (2010).
\bibitem{KLOEreview}
      KLOE Collaboration, F. Bossi  {\it et al.},
      Riv. Nuovo Cim. {\bf 031} 531 (2008),  and references therein. 
\bibitem{KLOE-BS}
      The KLOE collaboration, G.~D'Ambrosio and G.~Isidori,
     JHEP {\bf 0612} 011 (2006).
\bibitem{KLOE-CPT}
     M.~Palutan, 
     Flavianet Kaon Workshop, Capri, Italy (2008).
\bibitem{KLOE-int}
      KLOE Collaboration, F.~Ambrosino {\it et al.},
         Phys. Lett. {\bf B642} 315 (2006).
\bibitem{KLOE-dec}
      A.~Di Domenico and the KLOE collaboration, 
         J. Phys.: Conf. Ser. {\bf 171} 012008 (2009).
\bibitem{CPTQM}
      A.~Di Domenico, 
      PoS KAON09 038 (2009), and references therein. 
\bibitem{KLOE-2}
        {\tt   http://www.lnf.infn.it/kloe2/}; 
         P.~Branchini,
        PoS KAON09 048 (2009); 
        G. Amelino-Camelia {\it et al.},
        arXiv:1003.3868 (2010).
\bibitem{KLOE2-GEM}
      KLOE-2 Collaboration, G.~De Robertis  {\it et al.},
      arXiv:1002.2572 (2010).
\bibitem{Lunghi-Soni08}
     E.~Lunghi and A.~Soni, 
         Phys. Lett. {\bf B666} 162 (2008).
\bibitem{Buras-Guadagnoli08}
    A.J.~Buras and D.~Guadagnoli,
    Phys. Rev. {\bf D78} 033005 (2008).
\bibitem{Prell}
    S.~Prell, 
        in these Proceedings. 
\bibitem{FlaviaNet}
   {\tt   http://www.lnf.infn.it/wg/vus/}.
\bibitem{FN-Vus}
      FlaviaNet Working Group on Kaon Decays, M.~Antonelli {\it et al.},
      arXiv:1005.2323 (2010).
\bibitem{KAON09}
   {\tt   http://kaon09.kek.jp/}.
\bibitem{KAON09Proc}
   {\tt   http://pos.sissa.it/cgi-bin/reader/conf.cgi?confid=83}. 
   


%
%
\end{thebibliography}
\end{document}